\documentclass[american,english,conference]{IEEEtran}
\usepackage[T1]{fontenc}
\usepackage[latin9]{inputenc}
\usepackage{array}
\usepackage{verbatim}
\usepackage{float}
\usepackage{amsmath}
\usepackage{amsthm}
\usepackage{amssymb}
\usepackage{graphicx}

\makeatletter

\providecommand{\tabularnewline}{\\}
\floatstyle{ruled}
\newfloat{algorithm}{tbp}{loa}
\providecommand{\algorithmname}{Algorithm}
\floatname{algorithm}{\protect\algorithmname}

\theoremstyle{plain}
\newtheorem{thm}{\protect\theoremname}
\theoremstyle{plain}
\newtheorem{lem}[thm]{\protect\lemmaname}

\usepackage{algorithmic}
\usepackage{algorithm}
\usepackage{diagbox}

\bstctlcite{IEEEexample:BSTcontrol}
\usepackage{multibib}
\makeatletter
\def\bstctlcite{\@ifnextchar[{\@bstctlcite}{\@bstctlcite[@auxout]}}
\def\bstctlcite[#1]#2{\@bsphack
\@for\@citeb:=#2\do{%
\edef\@citeb{\expandafter\@firstofone\@citeb}%
\if@filesw\immediate\write\csname #1\endcsname{\string\citation{\@citeb}}\fi}%
\@esphack}
\makeatother
\newcites{sec}{Secondary Literature}
\bstctlcite[@auxoutsec]{BSTcontrol2}

\floatname{algorithm}{Algorithm}

\DeclareMathOperator{\tr}{tr}

\DeclareMathOperator{\diag}{diag}
\DeclareMathOperator{\maximize}{maximize}
\DeclareMathOperator{\st}{subject~to}
\newcommand{\herm}{^{{\dagger}}}
\newcommand{\trans}{^{\mathsf{T}}}

\makeatother

\usepackage{babel}
\addto\captionsamerican{\renewcommand{\algorithmname}{Algorithm}}
\addto\captionsamerican{\renewcommand{\lemmaname}{Lemma}}
\addto\captionsamerican{\renewcommand{\theoremname}{Theorem}}
\addto\captionsenglish{\renewcommand{\lemmaname}{Lemma}}
\addto\captionsenglish{\renewcommand{\theoremname}{Theorem}}
\providecommand{\lemmaname}{Lemma}
\providecommand{\theoremname}{Theorem}

\begin{document}
\title{Efficient Numerical Methods for Secrecy Capacity of Gaussian MIMO
Wiretap Channel}
\author{\IEEEauthorblockN{Anshu~Mukherjee\IEEEauthorrefmark{1}, Björn~Ottersten\IEEEauthorrefmark{2},
and~Le~Nam~Tran\IEEEauthorrefmark{1}}\IEEEauthorblockA{\IEEEauthorrefmark{1}School of Electrical and Electronic Engineering,
University College Dublin, Ireland\\
 Email: anshu.mukherjee@ucdconnect.ie; nam.tran@ucd.ie}\IEEEauthorblockA{\IEEEauthorrefmark{2}Interdisciplinary Centre for Security, Reliability
and Trust, University of Luxembourg, Luxembourg\\
 Email: bjorn.ottersten@uni.lu}}
\maketitle
\begin{abstract}
This paper presents two different low-complexity methods for obtaining
the secrecy capacity of multiple-input multiple-output (MIMO) wiretap
channel subject to a sum power constraint (SPC). The challenges in
deriving computationally efficient solutions to the secrecy capacity
problem are due to the fact that the secrecy rate is a difference
of convex functions (DC) of the transmit covariance matrix, for which
its convexity is only known for \emph{the degraded case}. In the first
method, we capitalize on the accelerated DC algorithm, which requires
solving a sequence of convex subproblems. In particular, we show that
each subproblem indeed admits a water-filling solution. In the second
method, based on the equivalent convex-concave reformulation of the
secrecy capacity problem, we develop a so-called partial best response
algorithm (PBRA). Each iteration of the PBRA is also done in closed
form. Simulation results are provided to demonstrate the superior
performance of the proposed methods.

\textit{Index Terms} - MIMO, secrecy capacity, sum power constraint,
convex-concave, closed form solution.
\end{abstract}

\section{Introduction:}

Wireless communication is an integral part of our modern life. Due
to its broadcasting nature, information sent over wireless channels
is vulnerable to security breach. Significant measures and techniques
have been developed by both industry and academia to address this
critical issue. Particularly, cryptography is a conventional method
to ensure data security in wireless networks. In recent years, physical
layer security has received growing interest as a promising alternative
to addressing wireless security. While cryptographic methods are based
on computational complexity and implemented in high network layers,
physical layer security is concerned with exploiting distinguishing
properties of wireless channels to achieve secure communication.

The wiretap channel (WTC), in which an eavesdropper aims to decode
to the message exchanged by a pair of legitimate transceivers, represents
a fundamental information-theoretic model for physical layer security.
The secrecy capacity of the WTC was first studied by Wyner in \cite{Wyner75}.
Since Wyner's seminal paper, the WTC has been extended, covering various
scenarios. In particular, the secrecy capacity of the Gaussian WTC
was studied \cite{Gauss_wiretap}. The use of multiple antennas at
transceivers in contemporary wireless communications systems naturally
gives rise to the so-called multiple-input multiple-output (MIMO)
Gaussian WTC. The secrecy capacity of Gaussian MIMO WTC has received
significant interests since the late 2000s. In this regard, there
have been many significant results in the literature, which are discussed
as follows.

The analytical solution for the Gaussian multiple-input single-out
(MISO) WTC where both the eavesdropper and the legitimate receiver
have a single antenna was proposed in \cite{ZangLi}. When the channel
state information is perfectly known, the secrecy capacity of MIMO
WTC was characterized in \cite{Secrecy_cap_MISOME,MIMOME_WTC,Oggier2011SecCapEq}.
Particularly, explicit expressions for optimal signaling for Gaussian
MIMO WTC are possible under some special cases \cite{Li2010MIMOwiretap,Sec_MIMO_SPC_3,Loyka2016}.
Power minimization and secrecy rate maximization for MIMO WTC was
studied in \cite{Cumanan2014} using a difference of convex functions
algorithm (DCA). More recently, a low-complex solution for Gaussian
MIMO WTC was proposed in \cite{ThangNguyen2020} using the equivalent
convex-concave reformulation of the secrecy capacity problem.%

In this paper we consider the problem of finding the secrecy capacity-achieving
input covariance for Gaussian MIMO WTC subject to a sum power constraint
(SPC). The case of general linear transmit covariance constraints
such as per-antenna power constraint is studied in \cite{Anshu:MIMOWTC:2020}.
In particular, we develop two low-complexity methods to solve the
secrecy-capacity problem for the Gaussian MIMO WTC. The first method
is an accelerated difference of convex functions algorithm (ADCA)
\cite{NhatPhan2017} where each subproblem is found in closed form.
In the second method, we propose an efficient iterative method to
calculate the secrecy capacity, which is based on the equivalent concave-convex
reformulation of the secrecy capacity problem. We refer to this proposed
method as the partial best response algorithm (PBRA). The idea of
PBRA is to find a saddle point of the concave-convex problem, for
which efficient numerical methods are also derived. We remark that
the method presented in \cite{ThangNguyen2020} is a double-loop iterative
algorithm, while the proposed PBRA in this paper only requires a single
loop.

\textit{Notation:} We use bold uppercase and lowercase letters to
denote matrices and vectors, respectively. $\mathbb{C}^{M\times N}$
denotes the space of $M\times N$ complex matrices.To lighten the
notation, $\mathbf{I}$ and $\mathbf{0}$ define identity and zero
matrices respectively, of which the size can be easily inferred from
the context. $\mathbf{H}\herm$ and $\mathbf{H}\trans$ are Hermitian
and ordinary transpose of $\mathbf{H}$, respectively; $\mathbf{H}_{i,j}$
is the $(i,j)$-entry of $\mathbf{H}$; $|\mathbf{H}|$ is the determinant
of $\mathbf{H}$; Furthermore, we denote the expected value of a random
variable by $\mathbb{E}[.]$. For $\mathbf{x}\in\mathbb{R}^{N}$ $[\mathbf{x}]_{+}=\begin{bmatrix}\max(x_{1},0), & \max(x_{2},0), & \cdots & \max(x_{N},0)\end{bmatrix}$.
The $i$th unit vector (i.e., its $i$th entry is equal to one and
all other entries are zero) is denoted by $\mathbf{e}_{i}$. The notation
$\mathbf{A}\succeq(\succ)\mathbf{B}$ means $\mathbf{A}-\mathbf{B}$
is positive semidefinite (definite). $\diag(\mathbf{x})$ creates
a diagonal matrix whose diagonal elements are taken from $\mathbf{x}$.
$\left\Vert \mathbf{H}\right\Vert $denotes the Frobenius norm of
$\mathbf{H}$.

\section{System Model}

\subsection{MIMO Wiretap Channel Model}

We consider a MIMO WTC, where Alice wants to transmit information
to the legitimate receiver Bob in presence of Eve, the eavesdropper.
The number of antennas at Bob, Alice and Eve is denoted by $N_{t}$,
$N_{r}$, and $N_{e}$, respectively. $\mathbf{H}_{b}\in\mathbb{C}^{N_{r}\times N_{t}}$
is the channel matrix between Alice and Bob. The channel matrix between
Bob and Eve is denoted by $\mathbf{H}_{e}\in\mathbb{C}^{N_{e}\times N_{t}}$.The
received signals at Bob and Eve are respectively expressed as \begin{subequations}
\begin{equation}
\mathbf{y}_{b}=\mathbf{H}_{b}\mathbf{x}+\mathbf{z}_{b}
\end{equation}
\begin{equation}
\mathbf{y}_{e}=\mathbf{H}_{e}\mathbf{x}+\mathbf{z}_{e}
\end{equation}
\end{subequations}where, $\mathbf{x}\in\mathbb{C}^{N_{t}\times1}$
represents the transmitted signal ; $\mathbf{z}_{b}\in\mathbb{C}^{N_{r}\times1}\sim\mathcal{CN}(\mathbf{0,I})$
and $\mathbf{z}_{e}\in\mathbb{C}^{N_{e}\times1}\sim\mathcal{CN}(\mathbf{0,I})$
are the additive white Gaussian noise at Bob and Eve, respectively.
\foreignlanguage{american}{I}n this paper $\mathbf{H}_{b}$ and $\mathbf{H}_{e}$
are assumed to be quasi-static and perfectly known at Alice and Bob.
For a given input covariance matrix $\mathbf{X}=E\{\mathbf{x}\mathbf{x}\herm\}\succeq\mathbf{0}$,
where $E\{\cdot\}$ is the statistical expectation, the maximum secrecy
rate (in nat/s/Hz) between Alice and Bob is given by \cite{Oggier2011SecCapEq}
\begin{equation}
C_{s}(\mathbf{X})=\left[\underbrace{\ln|\mathbf{I}+\mathbf{H}_{b}\mathbf{X}\mathbf{H}_{b}\herm|}_{f_{b}(\mathbf{X})}-\underbrace{\ln|\mathbf{I}+\mathbf{H}_{e}\mathbf{X}\mathbf{H}_{e}\herm|}_{f_{e}(\mathbf{X})}\right]_{+}.\label{SecrecyCapacityEqn}
\end{equation}
The secrecy capacity under the sum power constraint (SPC) is written
as\begin{subequations}\label{eq:secrecycapacity:org}
\begin{align}
\underset{\mathbf{X}\succeq\mathbf{0}}{\mathrm{maximize}} & \ C_{s}(\mathbf{X})\\
\st & \ \tr(\mathbf{X})\leq P_{0}
\end{align}
\end{subequations}where $P_{0}$ is the total transmit power. We
remark that the problem (\ref{eq:secrecycapacity:org}) is non-convex
in general, and thus, it is very difficult to find a globally optimal
solution. However, if the channel is degraded, i.e. $\mathbf{H}_{b}\herm\mathbf{H}_{b}\succeq\mathbf{H}_{e}\herm\mathbf{H}_{e}$,
(\ref{eq:secrecycapacity:org}) becomes convex but off-the-shelf solvers
cannot be used to solve it. In this regard, the equivalent minimax
reformulation of (\ref{eq:secrecycapacity:org}) presented in the
next subsection is more numerically useful.

\subsection{Minimax Reformulation}

It is interesting to note that the secrecy capacity of MIMO WTC in
(\ref{SecrecyCapacityEqn}) is equivalent to the following minimax
optimization problem 
\begin{equation}
C_{s}=\underset{\mathbf{Q}\in\mathcal{Q}}{\min}\ \underset{\mathbf{X}\in\mathcal{X}}{\max}\ f(\mathbf{Q},\mathbf{X})\triangleq\log\frac{|\mathbf{I}+\mathbf{Q}^{-1}\bar{\mathbf{H}}\mathbf{X}\bar{\mathbf{H}}\herm|}{|\mathbf{I}+\mathbf{H}_{e}\mathbf{X}\mathbf{H}_{e}\herm|}\label{eq:MiniMax}
\end{equation}
where $\bar{\mathbf{H}}=[\mathbf{H}_{b}\trans,\mathbf{H}_{e}\trans]\trans$
and $\mathbf{Q}\in\mathbb{C}^{N_{R}\times N_{E}}$. The sets $\mathcal{Q}$
and $\mathcal{X}$ are defined as 
\begin{equation}
\mathcal{X}=\{\mathbf{X}|\mathbf{X}\succeq\mathbf{0};\tr(\mathbf{X})=P_{0}\}
\end{equation}
and
\begin{equation}
\mathcal{Q}=\left\{ \mathbf{Q}|\mathbf{Q}\succeq\mathbf{0};\mathbf{Q}=\small\left[\begin{array}{cc}
\mathbf{I} & \bar{\mathbf{Q}}\\
\bar{\mathbf{Q}}\herm & \mathbf{I}
\end{array}\right]\right\} .
\end{equation}
Compared to (\ref{eq:secrecycapacity:org}), (\ref{eq:MiniMax}) is
more tractable since the objective of (\ref{eq:MiniMax}) is concave
with $\mathbf{X}$ for a given $\mathbf{Q}$ and convex with $\mathbf{Q}$
for a given $\mathbf{X}$. In particular, we can compute the secrecy
capacity and the optimal signaling by finding the saddle point of
$f(\mathbf{Q},\mathbf{X})$.

\section{Proposed Algorithms}

In this section we present two low-complexity methods for finding
the secrecy capacity of the MIMO WTC. The first method is a result
of applying an ADCA to (\ref{eq:secrecycapacity:org}) and the second
one is based on a finding a saddle point of (\ref{eq:MiniMax}).

\subsection{ADCA for Solving (\ref{eq:secrecycapacity:org})}

To solve (\ref{eq:secrecycapacity:org}), we propose a simple but
efficient method derived based on the obvious observation that $C_{s}(\mathbf{X})$
is a DC function, which naturally motivates the use of DCA. In this
work we apply the ADCA presented in \cite{NhatPhan2017}. The idea
is that from the current and previous iterates, denoted by $\mathbf{X}_{n}$
and $\mathbf{X}_{n-1}$ respectively, we compute an extrapolated point
$\mathbf{Z}_{n}$ using the Nesterov\textquoteright s acceleration
technique: $\mathbf{X}_{n}+(t_{k}-1)/t_{k+1}\bigl(\mathbf{X}_{n}-\mathbf{X}_{n-1}\bigr)$,
where $t$ is the acceleration parameter. We remark that the specific
$t$-update in Line \ref{alg:ADCA:t} is a condition to guarantee
the convergence of the iterative process as described in \cite{Nesterov1983AMF}.
Since $C_{s}(\mathbf{X})$ is possibly non-convex for a general MIMO
WTC, $\mathbf{Z}_{n}$ can be a bad extrapolation and a monitor is
required. Specifically, if $\mathbf{Z}_{n}$ is better than \emph{one
of the last} $q$ iterates, then $\mathbf{Z}_{n}$ is considered a
good extrapolation and thus will be used instead of $\mathbf{X}_{n}$
to generate the next iterate. Thus, the ADCA is generally \emph{non-monotone}.
The algorithmic description of ADCA for solving (\ref{eq:secrecycapacity:org})
is outlined in Algorithm \ref{alg:ADCA}. Note that the subproblem
in (\ref{eq:DCA:subprob}) is achieved by linearizing $f_{e}(\mathbf{X})$
around $\mathbf{V}_{n}$ and by omitting the associated constants
that do not affect the optimization. In Algorithm \ref{alg:ADCA},
$q$ is any non-negative integer and $\gamma_{n}$ is the minimum
of the secrecy rate of the last $q$ iterates. We remark that the
case when $q=0$ reduces to the conventional DCA, which is exactly
the same as the AO method in \cite{Li2013}.
\begin{algorithm}
\caption{ADCA for solving (\ref{eq:secrecycapacity:org})}

\label{alg:ADCA}

\begin{algorithmic}[1]

\STATE Initialization: $\mathbf{W}_{0}=\mathbf{X}_{0}\in\mathcal{X}$,
$t=\frac{1+\sqrt{5}}{2}$, $q$: integer.

\FOR{ $n=1,2,\dots$}

\STATE Update:
\begin{equation}
\mathbf{X}_{n}=\underset{\mathbf{X}\in\mathcal{X}}{\arg\max}\ \underbrace{f_{b}(\mathbf{X})-\tr\bigl(\nabla f_{e}(\mathbf{W}_{n-1})\mathbf{X}\bigr)}_{\bar{f}(\mathbf{X};\mathbf{W}_{n-1})}\label{eq:DCA:subprob}
\end{equation}
where $\nabla f_{e}(\mathbf{X})=\mathbf{H}_{e}\herm\bigl(\mathbf{I}+\mathbf{H}_{e}\mathbf{X}\mathbf{H}_{e}\herm\bigr)^{-1}\mathbf{H}_{e}$

\STATE $t_{n+1}=\frac{1+\sqrt{1+4t_{n}^{2}}}{2}$\label{alg:ADCA:t}

\STATE $\mathbf{Z}_{n}=\mathbf{X}_{n}+\frac{t_{n}-1}{t_{n+1}}\bigl(\mathbf{X}_{n}-\mathbf{X}_{n-1}\bigr)$

\STATE $\gamma_{n}=\min\bigl(C_{s}(\mathbf{X}_{n}),C_{s}(\mathbf{X}_{n-1}),\ldots,C(\mathbf{X}_{[n-q]_{+}})\bigr)$\label{alg:ADCA:gamma}

\STATE $\mathbf{W}_{n}=\begin{cases}
\mathbf{Z}_{n} & \textrm{if}\ C_{s}(\mathbf{Z}_{n})\geq\gamma_{n}\ \text{and}\ \mathbf{Z}_{n}\ \text{is feasible}\\
\mathbf{X}_{n} & \textrm{otherwise}
\end{cases}$

\ENDFOR

\STATE Output: $\mathbf{X}_{n}$

\end{algorithmic}
\end{algorithm}

To implement Algorithm \ref{alg:ADCA}, we need to efficiently solve
(\ref{eq:DCA:subprob}), which is explicitly written as \begin{subequations}\label{eq:MaxDCASubprob}
\begin{align}
\underset{\mathbf{X}\succeq\mathbf{0}}{\mathrm{maximize}} & \ \ln\bigl|\mathbf{I}+\mathbf{H}_{b}\mathbf{X}\mathbf{H}_{b}\herm\bigr|-\tr\bigl(\boldsymbol{\Phi}_{n-1}\mathbf{X}\bigr)\\
\st & \ \tr(\mathbf{X})=P_{0}
\end{align}
\end{subequations} where $\boldsymbol{\Phi}_{n-1}=\mathbf{H}_{e}\herm\bigl(\mathbf{I}+\mathbf{H}_{e}\mathbf{W}_{n-1}\mathbf{H}_{e}\herm\bigr)^{-1}\mathbf{H}_{e}$.
We now show that (\ref{eq:MaxDCASubprob}) admits a \emph{water-filling
solution} To begin with, let us form the \emph{partial} Lagrangian
function associated with (\ref{eq:DCA:subprob}) as
\begin{equation}
\mathcal{L}(\mathbf{X},\mu)=\ln|\mathbf{I}+\mathbf{H}_{b}\mathbf{X}\mathbf{H}_{b}\herm|-\tr\bigl(\bar{\boldsymbol{\Phi}}_{n-1}\mathbf{X}\bigr)+\mu P_{0}\label{eq:LagrangianDCAsubprob}
\end{equation}
where $\bar{\boldsymbol{\Phi}}_{n-1}=\boldsymbol{\Phi}_{n-1}+\mu\mathbf{I}_{N_{t}}$
and $\mu\geq0$ is the Lagrangian multiplier. Let $\bar{\mathbf{X}}\triangleq\bar{\boldsymbol{\Phi}}_{n-1}^{1/2}\mathbf{X}\bar{\boldsymbol{\Phi}}_{n-1}^{1/2}$
and rewrite the Lagrangian function as a function of $\bar{\mathbf{X}}$
as 
\begin{equation}
\mathcal{L}(\bar{\mathbf{X}},\mu)=\ln\bigl|\mathbf{I}+\mathbf{H}_{b}\bar{\boldsymbol{\Phi}}_{n-1}^{-1/2}\bar{\mathbf{X}}\bar{\boldsymbol{\Phi}}_{n-1}^{-1/2}\mathbf{H}_{b}\herm\bigr|-\tr(\bar{\mathbf{X}}).
\end{equation}
To derive the dual function, the following lemma is in order \cite{ThangNguyen2020}.
\begin{lem}
For a given $\mu\geq0$, let $\bar{\boldsymbol{\Phi}}_{n-1}^{-1/2}\mathbf{H}_{b}\herm\mathbf{H}_{b}\bar{\boldsymbol{\Phi}}_{n-1}^{-1/2}=\mathbf{V}\boldsymbol{\Sigma}\mathbf{V}\herm$
be the eigen value decomposition (EVD) of $\bar{\boldsymbol{\Phi}}_{n-1}^{-1/2}\mathbf{H}_{b}\herm\mathbf{H}_{b}\bar{\boldsymbol{\Phi}}_{n-1}^{-1/2}$,
where $\mathbf{V}\in\mathbb{C}^{N_{t}\times N_{t}}$ is unitary, $\boldsymbol{\Sigma}=\diag(\sigma_{1},\sigma_{2},\dots,\sigma_{r},0,\ldots,0)$,
and $r$ is the rank of $\bar{\boldsymbol{\Phi}}_{n-1}^{-1/2}\mathbf{H}_{b}$.
Then the solution to the problem $\underset{\bar{\mathbf{X}}\succeq\mathbf{0}}{\max}\mathcal{L}(\bar{\mathbf{X}},\mu)$
is given by
\begin{equation}
\mathbf{X}=\bar{\boldsymbol{\Phi}}_{n-1}^{-1/2}\mathbf{V}\bar{\boldsymbol{\Sigma}}\mathbf{V}\herm\bar{\boldsymbol{\Phi}}_{n-1}^{-1/2}\label{eq: XwaterfillSol}
\end{equation}
where $\mathbf{\bar{\boldsymbol{\Sigma}}=\diag}\bigl([1-\frac{1}{\sigma_{1}}]_{+},\ldots,[1-\frac{1}{\sigma_{r}}]_{+},0,\ldots,0\bigr)$.
\end{lem}
Next, to solve (\ref{eq:MaxDCASubprob}) we need to find the optimal
value of $\mu$ which can be done by a bisection search. We skip the
details here for the sake of brevity. The convergence proof of Algorithm
\ref{alg:ADCA} is provided in \cite{Anshu:MIMOWTC:2020}. The idea
is to show that the sequence $\{\gamma_{n}\}$ increasing and the
feasible set $\mathcal{X}$ is compact and convex. Thus, there exists
a convergent subsequence, the accumulation point of which is then
proved to be a stationary point.

\subsection{Partial Best Response Method for Solving (\ref{eq:MiniMax})}

The second proposed method is an iterative algorithm to find the saddle
point of the concave-convex problem in (\ref{eq:MiniMax}). Suppose
$(\mathbf{X}_{n-1},\mathbf{Q}_{n-1})$ has been computed at the $n$-th
iteration. Then $\mathbf{X}_{n}$ is found as 
\begin{align}
\mathbf{X}_{n} & =\underset{\mathbf{X}\in\mathcal{X}}{\arg\max}\ f(\mathbf{\mathbf{Q}}_{n},\mathbf{X})\nonumber \\
 & \hspace{-0.8cm}=\underset{\mathbf{X}\in\mathcal{X}}{\arg\max}\ \log|\mathbf{Q}_{n-1}+\mathbf{H}\mathbf{X}\mathbf{H}\herm|-\log|\mathbf{I}+\mathbf{H}_{e}\mathbf{X}\mathbf{H}_{e}\herm|.\label{eq:findX}
\end{align}
In words,\emph{ $\mathbf{X}_{k}$ is the best response to $\mathbf{\mathbf{Q}}_{n-1}$
as usual}. Now given $\mathbf{X}_{n}$, due to the concavity of the
term $\log|\mathbf{\mathbf{Q}}+\mathbf{H}\mathbf{X}\mathbf{H}\herm|$,
the following inequality holds
\begin{align}
f(\mathbf{\mathbf{Q}},\mathbf{X}_{n}) & \leq\log|\mathbf{\mathbf{Q}}_{n-1}+\mathbf{H}\mathbf{X}_{n}\mathbf{H}\herm|+\tr(\boldsymbol{\Psi}_{n}(\mathbf{\mathbf{Q}}-\mathbf{\mathbf{Q}}_{n-1}))\nonumber \\
 & \quad-\log(\mathbf{\mathbf{Q}})-\log|\mathbf{I}+\mathbf{H}_{e}\mathbf{X}_{n}\mathbf{H}_{e}\herm|,\forall\mathbf{\mathbf{\mathbf{Q}}}\in\mathcal{Q}.\nonumber \\
 & \triangleq\bar{f}(\mathbf{\mathbf{Q}},\mathbf{X}_{n}).\label{eq:findK:UB}
\end{align}
where $\boldsymbol{\Psi}_{n}=(\mathbf{\mathbf{Q}}_{n-1}+\mathbf{H}\mathbf{X}_{n}\mathbf{H}\herm)^{-1}$.
Note that the above inequality is tight when $\mathbf{\mathbf{Q}}=\mathbf{\mathbf{\mathbf{Q}}}_{n-1}$.
Next, $\mathbf{\mathbf{\mathbf{Q}}}_{n}$ is obtained as
\begin{equation}
\mathbf{\mathbf{Q}}_{n}=\underset{\mathbf{\mathbf{Q}}\in\mathcal{Q}}{\arg\min}\bar{f}(\mathbf{\mathbf{Q}},\mathbf{X}_{n})=\underset{\mathbf{\mathbf{Q}}\in\mathcal{Q}}{\arg\min}\tr(\boldsymbol{\Psi}_{n}\mathbf{\mathbf{Q}})-\log|\mathbf{\mathbf{Q}}|.\label{eq:findK}
\end{equation}
That is to say, $\mathbf{\mathbf{Q}}_{n}$ is found be the best response
to $\mathbf{X}_{n}$ using\emph{ an upper bound} of the objective.
The proposed solution for finding the secrecy capacity is summarized
in Algorithm \ref{alg:PABR}.
\begin{algorithm}
\caption{PBRA for solving (\ref{eq:MiniMax})}

\label{alg:PABR}

\begin{algorithmic}[1]

\STATE Input: $\mathbf{\mathbf{Q}}_{1}\ensuremath{\in}\mathcal{\mathcal{Q}}$,
$\epsilon_{1}>0$

\FOR{ $n=1,2\dots$}

\STATE Update $\mathbf{X}_{n}$ according to (\ref{eq:findX})\label{alg:PABR:X-update}

\STATE Update $\mathbf{\mathbf{Q}}_{n+1}$ according to (\ref{eq:findK})\label{alg:PABR:Q-update}

\ENDFOR

\STATE Output: $\mathbf{X}_{n}$

\end{algorithmic}
\end{algorithm}

We remark that the iterative method presented in \cite{ThangNguyen2020}
also aims to find the saddle-point of (\ref{eq:MiniMax}). However,
it is a double-loop algorithm where a lower bound of $f(\mathbf{\mathbf{Q}}_{n},\mathbf{X})$
is used for the $\mathbf{X}$-update. In contrast, Algorithm \ref{alg:PABR}
is a single-loop one where the $\mathbf{X}$-update is exact. The
efficient methods for the $\mathbf{X}$-update and $\mathbf{\mathbf{Q}}$-update
are detailed in the following subsections. 

\subsubsection{$\mathbf{X}$-update}

 To compute $\mathbf{X}_{n}$ as in Line \ref{alg:PABR:X-update}
of Algorithm \ref{alg:PABR}, we need to solve (\ref{eq:findX}) which
is a convex problem. Since the projection onto $\mathcal{X}$ can
be done in closed form, we can apply an accelerated projected gradient
method (APGM) \cite{Beck2009APG} to solve it efficiently, which is
described as follows. To avoid confusion we use the superscript to
denote the iteration count of the APGM. Suppose $\mathbf{Y}^{(k)}$,
the extrapolated point at iteration $k$, is available. The next iterate
$\mathbf{X}^{(k)}$ is found as
\begin{equation}
\mathbf{X}^{(k)}=p_{\mathcal{\mathcal{X}}}\Bigl(\mathbf{Y}^{(k)}+\frac{1}{\beta}\nabla f\bigl(\mathbf{\mathbf{Q}}_{n-1},\mathbf{Y}^{(k)}\bigr)\Bigr)
\end{equation}
where $\frac{1}{\beta}$ is a step size and $p_{\mathcal{\mathcal{X}}}(\bar{\mathbf{X}})$
denotes the projection of $\bar{\mathbf{X}}$ onto $\mathcal{X}$.
The gradient of $f(\mathbf{\mathbf{Q}}_{n-1},\mathbf{X})$ is given
by
\begin{align}
\nabla f\bigl(\mathbf{\mathbf{Q}}_{n-1},\mathbf{X}\bigr) & =\bigl(\mathbf{H}\herm\bigl(\mathbf{\mathbf{Q}}_{n-1}+\mathbf{H}\mathbf{X}\mathbf{H}\herm\bigr)^{-1}\mathbf{H}\bigr)-\nonumber \\
 & \quad\qquad\bigl(\mathbf{H}_{e}\herm\bigl(\mathbf{I}+\mathbf{H}_{e}\mathbf{X}\mathbf{H}_{e}\herm\bigr)^{-1}\mathbf{H}_{e}\bigr).
\end{align}
For a given point $\bar{\mathbf{X}}$, the projection $p_{\mathcal{\mathcal{X}}}(\bar{\mathbf{X}})$
is mathematically stated as 
\begin{equation}
\underset{\mathbf{X}\succeq\mathbf{0}}{\maximize}\left\{ \bigl\Vert\mathbf{X}-\bar{\mathbf{X}}\bigr\Vert\ \bigl|\ \tr(\mathbf{X})=P_{0}\right\} \label{eq:SPCproj}
\end{equation}
which admits a closed-form solution as 
\begin{equation}
\mathbf{X}=\mathbf{U}\diag\bigl(\bigl[\bigl[\bar{\boldsymbol{\sigma}}\bigr]_{+}-\tau\bigr]_{+}\bigr)\mathbf{U}\herm
\end{equation}
where $\bar{\mathbf{X}}=\mathbf{U}\diag(\bar{\boldsymbol{\sigma}})\mathbf{U}\herm$
be the eigenvalue decomposition of $\bar{\mathbf{X}}$, $\bar{\boldsymbol{\sigma}}=[\bar{\sigma}_{1},\bar{\sigma}_{2},\ldots,\bar{\sigma}_{r}]$
where $r$ is the rank of $\bar{\mathbf{X}}$, and $\tau$ is the
unique number such that $\sum_{i=1}^{r}\max(\bigl[\bar{\sigma}_{i}\bigr]_{+}-\tau,0)=P_{0}$.

The APGM for solving (\ref{eq:findX}) is outlined in Algorithm \ref{alg:AGPFindX}.
Note that a proper step size can be found by a backtracking line search
as done in Lines (\ref{alg:APMG:backtrack:start})-(\ref{alg:APMG:backtrack:end}).Starting
from the step size of the previous iteration, the idea of the backtracking
line search is to reduce it by a factor of $\theta$ until (\ref{alg:APMG:backtrack:end})
is met. That is, we try to find a lower quadratic approximation of
the objective at the current iterate. It is shown in \cite{Beck2009APG}
that \ref{alg:AGPFindX} achieves the optimal convergence rate of
$\mathcal{O}(1/k^{2})$.
\begin{algorithm}
\caption{Accelerated projected gradient method for solving (\ref{eq:findX})}

\label{alg:AGPFindX}

\begin{algorithmic}[1]

\STATE Input: $\mathbf{Y}^{(1)}=\mathbf{X}^{(0)}=\mathbf{X}_{n-1}$,
$\eta_{0}>0$, $\theta>1$, $\xi_{1}=1$.

\FOR{ $k=1,2,\dots$}

\STATE $\beta=\eta_{k-1}/\theta$

\REPEAT\label{alg:APMG:backtrack:start}

\STATE $\beta\leftarrow\theta\beta$

\STATE $\mathbf{X}^{(k)}=p_{\mathcal{\mathcal{X}}}\bigl(\mathbf{Y}^{(k)}+\frac{1}{\beta}\nabla f\bigl(\mathbf{\mathbf{Q}}_{n-1},\mathbf{Y}^{(k)}\bigr)\bigr)$

\UNTIL{ $f(\mathbf{\mathbf{Q}}_{n-1},\mathbf{X}^{(k)})\geq f(\mathbf{\mathbf{Q}}_{n-1},\mathbf{Y}^{(k)})+\bigl\langle\nabla f(\mathbf{\mathbf{Q}}_{n-1},\mathbf{Y}^{(k)}),(\mathbf{X}^{(k)}-\mathbf{Y}^{(k)})\bigr\rangle-\frac{\beta}{2}\bigl\Vert\mathbf{X}^{(k)}-\mathbf{Y}^{(k)}\bigr\Vert^{2}$
}\label{alg:APMG:backtrack:end}

\STATE $\xi_{k+1}=0.5(1+\sqrt{1+4\xi_{k}^{2}})$; $\eta_{k}=\beta$

\STATE $\mathbf{Y}^{(k+1)}=\mathbf{X}^{(k)}+\frac{\xi_{k}-1}{\xi_{k+1}}\bigl(\mathbf{X}^{(k)}-\mathbf{X}^{(k-1)}\bigr)$

\ENDFOR

\end{algorithmic}
\end{algorithm}

\subsubsection{$\mathbf{\mathbf{Q}}$-update}

A closed-form solution is also possible for the $\mathbf{\mathbf{Q}}$-update.
Specifically,we can partition $\boldsymbol{\Psi}_{n}$ into\foreignlanguage{american}{
\begin{equation}
\boldsymbol{\Psi}_{n}=\begin{bmatrix}\boldsymbol{\Psi}_{n,11} & \boldsymbol{\Psi}_{n,12}\\
\boldsymbol{\Psi}_{n,12}^{H} & \boldsymbol{\Psi}_{n,22}
\end{bmatrix}.
\end{equation}
}To lighten the notation, we will drop the subscript $n$ onwards.
Now, let $\boldsymbol{\Psi}_{12}\boldsymbol{\Psi}_{12}\herm=\mathbf{U}_{\mathbf{\boldsymbol{\Psi}}}\bar{\boldsymbol{\Sigma}}_{\mathbf{\mathbf{\boldsymbol{\Psi}}}}\mathbf{U}_{\mathbf{\boldsymbol{\Psi}}}^{\dagger}$
be the eigenvalue decomposition of $\boldsymbol{\Psi}_{12}\boldsymbol{\Psi}_{12}\herm$
and $\bar{\boldsymbol{\Sigma}}_{\mathbf{\mathbf{\boldsymbol{\Psi}}}}=\diag(\sigma_{\boldsymbol{\Psi}_{1}},\sigma_{\boldsymbol{\Psi}_{2}},\ldots,\sigma_{\boldsymbol{\Psi}_{N_{r}}})$.
Then the optimal solution to (\ref{eq:findK}) is given by
\begin{equation}
\bar{\mathbf{Q}}_{n+1}=-\mathbf{U}_{\mathbf{\boldsymbol{\Psi}}}\Xi\mathbf{U}_{\mathbf{\boldsymbol{\Psi}}}\herm\boldsymbol{\Psi}_{12}\label{eq:Qupdate}
\end{equation}
where 
\begin{gather}
\Xi_{\boldsymbol{\Psi}}=2\diag\Bigl(\frac{1}{1+\sqrt{1+4\sigma_{\boldsymbol{\Psi}_{1}}}},\frac{1}{1+\sqrt{1+4\sigma_{\boldsymbol{\Psi}_{2}}}},\ldots,\nonumber \\
\frac{1}{1+\sqrt{1+4\sigma_{\boldsymbol{\Psi}_{N_{r}}}}}\Bigr).
\end{gather}
We refer the reader to \cite{Anshu:MIMOWTC:2020} for the proof of
(\ref{eq:Qupdate}).

The main idea behind the convergence proof of Algorithm \ref{alg:PABR}
is show the monotonic decrease of the objective sequence $f(\mathbf{\mathbf{Q}}_{n},\mathbf{X}_{n})$,
which is due to the fact that the term $\log|\mathbf{Q}+\mathbf{H}\mathbf{X}\mathbf{H}\herm|-\log|\mathbf{I}+\mathbf{H}_{e}\mathbf{X}\mathbf{H}_{e}\herm|$
is jointly concave with $\mathbf{Q}$ and $\mathbf{X}$. We refer
the interested reader to \cite{Anshu:MIMOWTC:2020} for further details.

\section{Numerical Results}

In this section we provide numerical results to evaluate the proposed
algorithms. We adopt the Kronecker model in our numerical investigation
\cite{Chuah2002}. Specifically, the channel between Alice and Bob
$\mathbf{H}_{b}$ is modeled as $\mathbf{H}_{b}=\tilde{\mathbf{H}}_{b}\mathbf{R}_{b}^{1/2}$,
where $\tilde{\mathbf{H}}_{b}$ is a matrix of i.i.d. complex Gaussian
distribution with zero mean and unit variance and $\mathbf{R}_{b}^{1/2}$
the corresponding a transmit correlation matrix. Here we adopt the
exponential correlation model whereby $[\mathbf{R}_{b}]_{i,j}=\bigl(re^{j\phi_{b}}\bigr){}^{|i-j|}$
for a given $r\in[0,1]$ and $\phi_{b}\in[0,2\pi)$. The channel between
Alice and Eve is modeled as $\mathbf{H}_{e}=\gamma\tilde{\mathbf{H}}_{e}\mathbf{R}_{e}^{1/2}$
for a given $\gamma>0$ and $\tilde{\mathbf{H}}_{e}$ and $\mathbf{R}_{e}$
are generated in the same way. The purpose of introducing $\gamma$
is to study the secrecy capacity of the MIMO WTC with respect to the
relative average strength of $\mathbf{H}_{b}$ and $\mathbf{H}_{e}$.
For the simulation purpose we use $\phi_{e}=\pi/2$, $\gamma=0.9$
and $r=0.9$. The codes of all algorithms in comparison were written
in MATLAB and executed in a 64-bit Windows PC with 16GB RAM and Intel
Core i7, 3.20 GHz. Note that since the noise power is normalized to
unity and thus $P_{0}$ is defined to be the signal to noise ratio
(SNR) in this section. In all simulations results, the parameter $q$
for Algorithm \ref{alg:ADCA} is taken as $q=5$.

\begin{figure}
\centering\includegraphics[bb=62bp 559bp 285bp 741bp,width=0.78\columnwidth]{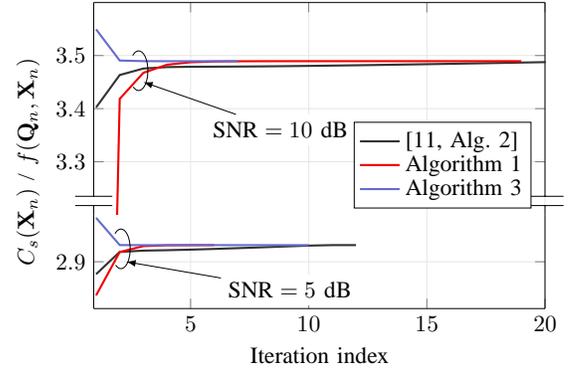}

\caption{Convergence results of iterative algorithms for different SNRs}

\label{fig:convergence}
\end{figure}

\begin{table}
\caption{Comparison of run-time (in milliseconds) between the proposed methods
and \cite[Alg. 2]{ThangNguyen2020}.}
\label{table: Table1}
\centering{}%
\begin{tabular}{c|>{\centering}p{0.8cm}|c|>{\centering}p{0.8cm}|c}
\hline 
 & \multicolumn{2}{c|}{$\begin{array}{c}
(N_{t},N_{r},N_{e})\\
=(4,3,2)
\end{array}$} & \multicolumn{2}{c}{$\begin{array}{c}
(N_{t},N_{r},N_{e})\\
=(4,6,8)
\end{array}$}\tabularnewline
\hline 
\diagbox{Algorithm}{SNR} & 5dB & 10dB & 5dB & 10dB\tabularnewline
\hline 
Algorithm \ref{alg:ADCA} & 14.6 & 17.2 & \textbf{26.8} & \textbf{41.5}\tabularnewline
\hline 
Algorithm \ref{alg:PABR} & \textbf{8.8} & \textbf{11.2} & 35.4 & 44.5\tabularnewline
\hline 
\cite[Alg. 2]{ThangNguyen2020} & 52.5 & 55.8 & 217.1 & 371.0\tabularnewline
\hline 
\end{tabular}
\end{table}

In Fig. \ref{fig:convergence} we show the convergence results of
proposed algorithms over two different SNRs $5$ dB and $10$ dB for
a set of randomly generated channel where $(N_{t},N_{r},N_{e})=(4,3,4)$.
For Algorithm \ref{alg:ADCA} we plot the secrecy rate $C_{s}(\mathbf{X}_{n})$
where $\mathbf{X}_{n}$ is the solution returned at the $n$th iteration.
For Algorithm \ref{alg:PABR} we plot the objective $f(\mathbf{Q}_{n},\mathbf{X}_{n})$
in (\ref{eq:MiniMax}). We also plot the convergence of the outer
loop of \cite[Algorithm 2]{ThangNguyen2020} for comparison. It is
clearly seen that the proposed algorithms converge very fast and all
algorithms converge to the same objective. For Algorithm \ref{alg:PABR},
we can also see that $f(\mathbf{Q}_{n},\mathbf{X}_{n})$ is indeed
an upper bound of $C_{s}(\mathbf{X}_{n})$ and it keeps decreasing
until convergence as expected. We remark that while Algorithm \ref{alg:ADCA}
is developed based on a local optimization method, our extensive numerical
results show that Algorithm \ref{alg:ADCA} always achieves the same
solution as Algorithm \ref{alg:PABR} which is optimal.

In Fig. \ref{fig:convergence} it also appears that all algorithms
in comparison achieve similar convergence rate performance in terms
of the required number of iterations. However, the complexity per
iteration of each algorithm is different. To achieve a more meaningful
comparison, we present their \emph{average actual run time} in Table
\ref{table: Table1}. For this purpose, the stopping criterion of
all algorithms is when the corresponding objective is not improved
during the last $5$ iterations. The average run time in Table \ref{table: Table1}
is obtained from 1000 random channel realizations. We can see that
the proposed algorithms, i.e. Algorithms \ref{alg:ADCA} and \ref{alg:PABR},
outperform \cite[Algorithm 2]{ThangNguyen2020}; and \ref{alg:PABR}
is slightly better than Algorithm \ref{alg:ADCA} when $N_{t}>N_{e}$
and vice versa when $N_{t}<N_{e}$. 
\begin{figure}[h]
\centering\includegraphics[bb=62bp 554bp 284bp 738bp,width=0.78\columnwidth]{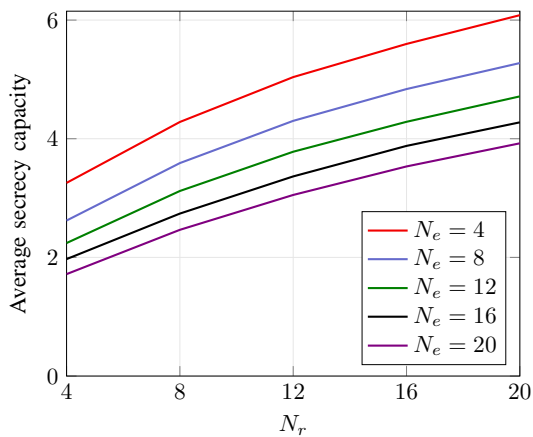}

\caption{Secrecy capacity as a function of $N_{r}$ for different values of
$N_{e}$. The number of transmit antennas is $N_{t}=4$.}

\label{fig:Tx_Impact}
\end{figure}

We now study how the secrecy capacity scales with the number of transmit
antennas at Bob and Alice. Fig. \ref{fig:Tx_Impact} plots the average
secrecy capacity for various numbers of antennas at Eve. The number
of transmit antennas at Alice is $N_{r}=4$. As can be seen in Fig.
\ref{fig:Tx_Impact}, the secrecy capacity increases with the number
of receive antennas at Alice, which is expected. Simultaneously, we
also observe that the secrecy capacity is reduced when the number
of antennas at Eve increases. In particular, Eve can significantly
decrease the secrecy capacity when $N_{e}$ is much larger than $N_{t}$.
This is because the null space of $\mathbf{H}_{b}$ will increasingly
intersect with the space spanned by $\mathbf{H}_{e}$. 

\begin{figure}
\centering\includegraphics[bb=62bp 556bp 294bp 741bp,width=0.78\columnwidth]{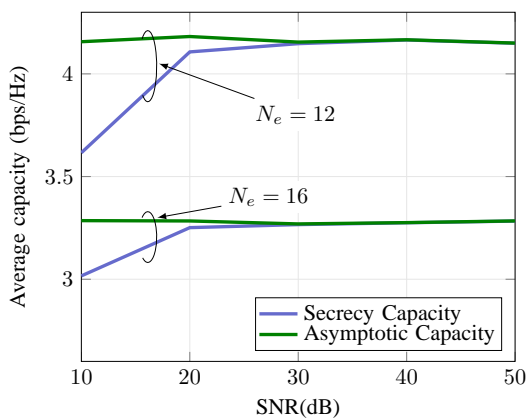}

\caption{Impact of $N_{e}$and SNR on the secrecy capacity and asymptotic capacity
at $N_{t}=6$, $N_{r}=4$}

\label{fig:AssympCap}
\end{figure}
Finally, Figure \ref{fig:AssympCap} plots the average secrecy capacity
as a function of SNR for different numbers of antennas at Eve. The
purpose is to understand the gap between the true secrecy capacity
and the asymptotic capacity obtained in \cite{MIMOME_WTC}. As expected,
the true secrecy capacity converges to the asymptotic capacity when
the SNR is sufficiently high. Again, we can observe the secrecy capacity
decreases when the number of receive antennas at Eve increases.

\section{Conclusion}

In this paper, we have proposed two efficient numerical methods for
computing the secrecy capacity and the optimal signaling of MIMO WTC.
In the first method, the secrecy capacity problem is viewed as a DC
program and we have applied an accelerated version of the celebrated
DCA, referred to as the ADCA. In the second method, we have drawn
on the convex-concave reformulation of the secrecy capacity problem
and developed the PBRA in which each iteration is done in closed form.
Numerical results have been provided to demonstrate that the proposed
solutions can reduce the run time of a known solution by 5 times for
the considered scenarios. Moreover, through extensive numerical experiments,
we have observed that the ADCA, albeit inherently a local optimization
method, always achieve the optimal solution. Our conjecture is that
the proposed ADCA is indeed a global optimization method, the proof
of which is left for future work.

\section*{Acknowledgment}

This publication has emanated from research supported by a Grant from
Science Foundation Ireland under Grant number 17/CDA/4786.

\bibliographystyle{IEEEtran}
\bibliography{IEEEabrv,paperVTC21}

\end{document}